\documentclass[aps,preprint]{revtex4}%
\usepackage{amsfonts}
\usepackage{amsmath}
\usepackage{amssymb}
\usepackage{graphicx}%
\providecommand{\U}[1]{\protect \rule{.1in}{.1in}}

\begin{document}
\title{Many-polaron description of impurities in a Bose-Einstein condensate in the
weak coupling regime}
\author{W. Casteels$^{1}$, J. Tempere$^{1,2}$ and J. T. Devreese$^{1}$}
\affiliation{$^{1}$TQC, Universiteit Antwerpen, Groenenborgerlaan 171, B2020 Antwerpen, Belgium}
\affiliation{$^{2}$Lyman Laboratory of Physics, Harvard University, Cambridge,
Massachusetts 02138, USA}

\begin{abstract}
The weak coupling many-polaron formalism is applied to the case of the
polaronic system consisting of impurities in a Bose-Einstein condensate. This
allows to investigate the groundstate properties and the response of the
system to Bragg spectroscopy. This theory is then applied to the system of
spin-polarized fermionic lithium-6 impurities in a sodium condensate. The
Bragg spectrum reveals a peak which corresponds to the emission of Bogoliubov
excitations. Both ground state properties and the response spectrum show that
the polaronic effect vanishes at large densities. We also look at two
possibilities to define the polaronic effective mass and observe that this
results in a different quantitative behavior if multiple impurities are involved.

\end{abstract}
\maketitle

\section{Introduction}

Quantum gases have revealed themselves as excellent quantum simulators for
many-body theories from condensed matter physics, and in particular to
experimentally examine strong coupling regimes that are not attainable with
solid state experiments\cite{RevModPhys.80.885}. Recently it was shown that
when the Bogoliubov approximation is valid the system of impurities in a
Bose-Einstein condensate (BEC) can be added to this list through a mapping
onto the Fr\"{o}hlich polaron
Hamiltonian\cite{PhysRevLett.96.210401,PhysRevA.73.063604}. The Fr\"{o}hlich
solid state polaron consists of a charge carrier (electron, hole) interacting
with the LO-phonons in an ionic crystal or a polar
semiconductor\cite{LandauPekar,Frolich3}. The Fr\"{o}hlich polaron Hamiltonian
has resisted an exact analytical diagonalization and has been submitted to
many approximation methods (a review on the Fr\"{o}hlich solid state polaron
can be found in Ref. \cite{Devreese}). The polaronic effect of an impurity in
a BEC has been the subject of several recent theoretical studies. Some
examples are the polaronic effects in optical
lattices\cite{PhysRevA.76.011605, 1367-2630-10-3-033015, PhysRevA.82.063614},
the application of the Feynman variational path integral technique to study
the ground state properties\cite{PhysRevB.80.184504}, an extension of this
technique to examine the response properties\cite{PhysRevA.83.033631} and a
strong coupling approximation\cite{springerlink:10.1134/S1054660X11150035}.
This revealed that the static properties are characterized by the ratio of the
masses of the bosons and the impurity together with the polaronic coupling
parameter $\alpha$, defined as:
\begin{equation}
\alpha=\frac{a_{IB}^{2}}{\xi a_{BB}}, \label{CouplPar}%
\end{equation}
with $a_{IB}\,$the impurity-boson scattering length, $a_{BB}$ the boson-boson
scattering length and $\xi$ the healing length of the condensate. As a
function of this coupling parameter two polaronic regimes were identified,
reminescent of the acoustic polaron\cite{JPSJ.35.137}. In the context of the
Fr\"{o}hlich solid state polaron the intermediate and strong coupling regimes
still supply important theoretical challenges but the weak coupling regime is
well understood. The weak and intermediate coupling theory of Lee, Low and
Pines\cite{PhysRev.90.297} was extended by Lemmens, Brosens and
Devreese\cite{GroundstateManyPolaron} to the case of an interacting polaron
gas. Similarly, the optical absorption theory of Devreese, Goovaerts and De
Sitter\cite{PhysRevB.5.2367} was extended to the case of interacting polarons
by two of the present authors, in Ref. \cite{PhysRevB.64.104504}. In these
extensions, the effect of the interactions between the polarons is taken into
account through the structure factor of the electron gas. The goal of this
paper is to apply a similar generalization to extend the polaronic theory of a
single impurity in a Bose condensate to the case of a dilute gas of
interacting impurities. This is needed since an experimental realization of
the BEC-impurity polaron will involve multiple impurties and will require the
use of Feshbach resonances to reach the strong coupling
regime\cite{PhysRevB.80.184504}.

Within the Bogoliubov approximation the Hamiltonian of $N$ impurities in a
condensate can be mapped onto the sum of the mean field energy $E_{MF}$ and
the Fr\"{o}hlich $N$-polaron Hamiltonian $\widehat{H}_{pol}^{N}$%
\cite{PhysRevB.80.184504}. The mean field energy is given by:%
\begin{equation}
E_{MF}=E_{GP}+N_{0}NU_{IB}\left(  \vec{q}=\vec{0}\right)  ,
\label{ExtraTermen}%
\end{equation}
where the first term is the Gross-Pitaevskii energy of the
condensate\cite{Pitaevskii} and the second term is the interaction shift due
to the impurities with $N_{0}$ the number of condensed bosons and
$U_{IB}\left(  \vec{q}\right)  $ the Fourier transform of the impurity-boson
interaction potential. The polaron Hamiltonian describes the mutual
interaction between the impurities and their interaction with the Bogoliubov
excitations:%
\begin{equation}
\widehat{H}_{pol}^{N}=\sum_{i}^{N}\frac{\widehat{\vec{p}}_{i}^{2}}{2m_{I}%
}+\sum_{\vec{k}}\hbar \omega_{\vec{k}}\widehat{a}_{\vec{k}}^{\dag}\widehat
{a}_{\vec{k}}+\sum_{\vec{k}}\sum_{i}^{N}\left(  V_{\vec{k}}\widehat{a}%
_{\vec{k}}e^{i\vec{k}.\widehat{\vec{r}}_{i}}+V_{\vec{k}}^{\dag}\widehat
{a}_{\vec{k}}^{\dag}e^{-i\vec{k}.\widehat{\vec{r}}_{i}}\right)  +\frac{1}%
{2}\sum_{i\neq j}v\left(  \widehat{\vec{r}}_{i}-\widehat{\vec{r}}_{j}\right)
. \label{NPolHam}%
\end{equation}
The first term represents the kinetic energy of the impurities with mass
$m_{I}$ and position (momentum) operators $\widehat{\vec{r}}_{i}$
($\widehat{\vec{p}}_{i}$), the second term gives the kinetic energy of the
Bogoliubov excitations with dispersion $\omega_{\vec{k}}$ and creation
(annihilation) operator $\widehat{a}_{\vec{k}}^{\dag}$ ($\widehat{a}_{\vec{k}%
})$, the third term is the interaction between the impurities and the
Bogoliubov excitations with interaction amplitude $V_{\vec{k}}$ and the fourth
term represents the mutual interaction between the impurities with $v\left(
\vec{r}\right)  $ the interaction potential. The Bogoliubov dispersion is
given by:%
\begin{equation}
\omega_{\vec{k}}=ck\sqrt{1+\left(  \xi k\right)  ^{2}/2}, \label{BogDisp}%
\end{equation}
where the speed of sound in the condensate was introduced: $c=\hbar/\left(
\sqrt{2}m_{B}\xi \right)  $, with $m_{B}$ the mass of the bosons. The
interaction amplitude is given by:%
\begin{equation}
V_{\vec{k}}=\sqrt{N_{0}}\left[  \frac{\left(  \xi k\right)  ^{2}}{\left(  \xi
k\right)  ^{2}+2}\right]  ^{1/4}g_{IB}, \label{IntAmp}%
\end{equation}
where a contact potential was assumed for the impurity-boson interaction
potential: $U_{IB}\left(  \vec{q}\right)  =g_{IB}$. At low temperature the
amplitude $g_{IB}$ is completely determined by the effective mass
$m_{r}=\left(  m_{B}^{-1}+m_{I}^{-1}\right)  ^{-1}$ and the impurity-boson
scattering length $a_{IB}$: $g_{IB}=2\pi \hbar^{2}a_{IB}/m_{r}$.

Another important consideration is the stability of the mixture against phase
separation. At zero temperature it was shown in Refs.
\cite{PhysRevA.61.053605} and \cite{PhysRevA.61.053601}\ that the mixture will
be stable if the following inequality is satisfied:%
\[
n^{1/3}\leq \frac{\left(  6\pi \right)  ^{2/3}}{12\pi}\frac{m_{r}a_{BB}}%
{m_{B}m_{I}a_{IB}^{2}},
\]
with $n$ the density of the impurities.

Since the impurities are not charged the response can not be studied through
optical absorption measurements as for the Fr\"{o}hlich solid state polaron.
In Ref. \cite{PhysRevA.83.033631} it was shown that the internal excitation
structure of the BEC-impurity polaron can be probed with Bragg spectroscopy
which is a technique that has proven to be very successful in the study of
Bose-Einstein condensates (see for example Refs. \cite{PhysRevLett.83.2876}
and \cite{PhysRevLett.88.120407}). The experimental setup consists of two
laser beams with wave vectors $\vec{k}_{1}$ and $\vec{k}_{2}$ and energies
$\omega_{1}$ and $\omega_{2}$ which are impinged on the impurities. These can
then absorb a photon from one beam and emit it in the other beam which results
in the exchange of a wave vector $\vec{k}=\vec{k}_{1}-\vec{k}_{2}$ and an
energy $\hbar \omega=\hbar \omega_{1}-\hbar \omega_{2}$ to the impurities. The
response of the system after an exposure during a time interval $\tau$ can be
measured by counting the number of atoms $N_{Bragg}$ that have gained a wave
vector $\vec{k}$ as a function of $\omega$ which in the formalism of linear
response theory is given by\cite{Pitaevskii}:%
\begin{equation}
N_{Bragg}=\frac{2}{\hbar}\left(  \frac{V}{2}\right)  ^{2}\tau \operatorname{Im}%
\chi \left(  \vec{k},\omega \right)  , \label{BraggResp}%
\end{equation}
with $V$ the amplitude of the laser induced potential and $\chi \left(  \vec
{k},\omega \right)  $ the density response function, defined as:%
\begin{equation}
\chi \left(  \vec{k},\omega \right)  =\frac{i}{\hbar}\int dte^{i\omega
t}\left \langle \rho_{\vec{k}}\left(  t\right)  \rho_{\vec{k}}^{\dag
}\right \rangle . \label{DensResp}%
\end{equation}

In the following we start by summarizing the main results of Refs.
\cite{GroundstateManyPolaron} and \cite{PhysRevB.64.104504} on the weak
coupling treatment of the Fr\"{o}hlich solid state $N$-polaron system are
summarized together with an indication of what changes if one considers
impurities in a BEC. We also show two different ways to define the polaronic
effective mass. This formalism is then applied for spin-polarized fermionic
impurities and the results are examined as a function of the impurity density
and the exchanged momentum for the Bragg response for lithium-6 impurities in
a sodium condensate.

\section{Weak coupling treatment of the many-polaron gas}

\subsection{Ground state properties of the many-polaron gas and the inertial
effective mass}

The weak coupling variational method as introduced in Ref.
\cite{PhysRev.90.297} by Lee, Low and Pines for the description of a single
polaron was generalized in Ref. \cite{GroundstateManyPolaron} to the case of a
many-polaron system. With this purpose the following variational wave function
was introduced:%
\begin{equation}
|\Psi_{LDB}>=\widehat{U}\left \vert 0\right \rangle \left \vert \psi \right \rangle
, \label{LDBWaveFun}%
\end{equation}
with $\left \vert 0\right \rangle $ the vacuum wave function for the bosonic
excitations, $\left \vert \psi \right \rangle $ the wave function of the
impurities and $\widehat{U}$ a canonical transformation of the form:%
\begin{equation}
\widehat{U}=\exp \left \{  \sum_{i}^{N}\sum_{\vec{k}}\left[  f_{\vec{k}}%
\widehat{a}_{\vec{k}}e^{i\vec{k}.\widehat{\vec{r}}_{i}}-f_{\vec{k}}^{\ast
}\widehat{a}_{\vec{k}}^{\dag}e^{-i\vec{k}.\widehat{\vec{r}}_{i}}\right]
\right \}  ,
\end{equation}
where $\left \{  f_{\vec{k}}\right \}  $ are variational functions. Minimizing
the expectation value of the $N$-polaron Hamiltonian (\ref{NPolHam}) with
respect to the variational wave function (\ref{LDBWaveFun})$\ $as a function
of $\left \{  f_{\vec{k}}\right \}  $ results in the following expression for
the ground state energy:%
\begin{align}
\frac{E}{N}  &  =\varepsilon_{kin}+\frac{1}{2}\sum_{\vec{k}}v\left(  \vec
{k}\right)  \left[  S\left(  \vec{k}\right)  -1\right] \nonumber \\
&  -\sum_{\vec{k}}\frac{S^{2}\left(  \vec{k}\right)  \left \vert V_{\vec{k}%
}\right \vert ^{2}}{\hbar \omega_{\vec{k}}S\left(  \vec{k}\right)  +\frac
{\hbar^{2}k^{2}}{2m_{I}}}, \label{GroundEnerg}%
\end{align}
where $\varepsilon_{kin}$ is the kinetic energy per particle: $\varepsilon
_{kin}=N^{-1}\left \langle \psi \left \vert \sum_{i}^{N}\frac{\widehat{p}_{i}%
^{2}}{2m_{I}}\right \vert \psi \right \rangle $ and $S\left(  \vec{k}\right)  $
is the static structure factor of the impurities:%
\begin{equation}
S\left(  \vec{k}\right)  =1+\frac{1}{N}\left \langle \psi \left \vert \sum_{i\neq
j}^{N}e^{i\vec{k}.\left(  \vec{r}_{i}-\vec{r}_{j}\right)  }\right \vert
\psi \right \rangle . \label{StatStruc}%
\end{equation}

Another important property of the polaronic system is the effective mass of
the polarons $m^{\ast}$. Within the one-polaron weak coupling formalism
$m^{\ast}$ was calculated in Ref. \cite{PhysRev.90.297}. The generalization to
many polarons allows a determination of this effective mass which is related
to the many-polaron system as a whole and shall be called the inertial
effective mass in the following. The total momentum $\mathcal{\vec{P}=}%
\sum_{i}\vec{p}_{i}+\sum_{\vec{k}}\hbar \vec{k}\widehat{a}_{\vec{k}}^{\dag
}\widehat{a}_{\vec{k}}$ commutes with the Hamiltonian (\ref{NPolHam}) and thus
is a conserved quantity. This conserved quantity can be explicitly introduced
in the minimization process by means of a Lagrange multiplier $\vec{v}$, which
corresponds to the velocity of the polaronic system:%
\begin{equation}
\widehat{H}_{pol}^{N}\left(  \vec{v}\right)  =\widehat{H}_{pol}^{N}-\vec
{v}.\left(  \sum_{i}\widehat{\vec{p}}_{i}+\sum_{\vec{k}}\hbar \vec{k}%
\widehat{a}_{\vec{k}}^{\dag}\widehat{a}_{\vec{k}}-\mathcal{\vec{P}}\right)  .
\label{Hamv}%
\end{equation}
minimization of the expectation value of (\ref{Hamv}) with respect to
(\ref{LDBWaveFun}) as a function of $\left \{  f_{\vec{k}}\right \}  $ and
$\vec{v}$, together with an Taylor expansion for small $\vec{v}$, results in:%
\begin{equation}
\mathcal{\vec{P}}=Nm_{I}\vec{v}+\frac{2}{3}\hbar^{2}N\sum_{\vec{k}}%
\frac{\left \vert V_{\vec{k}}\right \vert ^{2}S^{2}\left(  k\right)  }{\left(
\hbar \omega_{\vec{k}}S\left(  k\right)  +\frac{\hbar^{2}k^{2}}{2m_{I}}\right)
^{3}}k^{2}\vec{v},
\end{equation}
where isotropy of the system was assumed. We can now identify the inertial
effective mass $m^{\ast}$ through the expression $\mathcal{\vec{P}}=Nm^{\ast
}\vec{v}$, which relates the speed of the polaron to its momentum, i.e.:
\begin{equation}
m^{\ast}=m_{I}+\frac{2}{3}\hbar^{2}\sum_{\vec{k}}\frac{\left \vert V_{\vec{k}%
}\right \vert ^{2}S^{2}\left(  k\right)  }{\left(  \hbar \omega_{\vec{k}%
}S\left(  k\right)  +\frac{\hbar^{2}k^{2}}{2m_{I}}\right)  ^{3}}k^{2}.
\label{EffMass}%
\end{equation}
Note that the inertial effective mass (\ref{EffMass}) is a property of the
whole polaron system. This is important for an experiment where the behavior
of all the impurities together is studied. For example when one studies the
collective oscillation of the impurities in a harmonic trap the resulting
effective oscillation frequency is a function of this inertial effective mass.
This effect has been used to experimentally determine the effective mass of
Fermi polarons in Ref. \cite{PhysRevLett.103.170402}.

Note that for the ground state properties the presence of multiple impurities
is completely described through the static structure factor of the impurities.
In the limit of vanishing density the static structure factor becomes $1$ and
the expressions for the energy and the effective mass of a single polaron from
Ref. \cite{PhysRev.90.297} are retrieved.

\subsection{Bragg response of the many-polaron gas and the spectral effective
mass}

In Ref. \cite{PhysRevB.64.104504} the optical absorption of the $N$-polaron
gas consisting of electrons interacting with phonons was calculated within the
weak coupling formalism. The polaronic system consisting of impurities in a
BEC can not be probed with optical absorption but instead an appropriate
experimental technique is Bragg spectroscopy as shown in Ref.
\cite{PhysRevA.83.033631}. This means that a finite momentum exchange has to
be taken into account which is negligible in the case of the optical
absorption calculation. Furthermore it is the density-density correlation
function (\ref{DensResp}) that determines the response instead of the
current-current correlation which is needed for the optical absorption. These
two are however closely related since the application of two partial
integrations transforms equation (\ref{DensResp}) into a correlation function
of the time derivative of the density $\dot{\rho}_{\vec{k}}$ after which one
can use the Fourier transform of the continuity equation ($\dot{\rho}_{\vec
{k}}=i\vec{k}.\vec{j}_{\vec{k}}$) to obtain a current-current correlation
function:%
\begin{equation}
\chi \left(  \omega,\vec{k}\right)  =-\frac{i}{\hbar \omega^{2}}\int_{0}%
^{\infty}dte^{i\omega t}\left \langle \left[  \vec{k}.j_{\vec{k}}\left(
t\right)  ,\vec{k}.j_{\vec{k}}^{\dag}\right]  \right \rangle .
\label{DensResp2}%
\end{equation}
For the calculation of (\ref{DensResp2}) we use the derivation of Ref.
\cite{PhysRevB.64.104504} which is based on the wave function
(\ref{LDBWaveFun}) and which preserves only terms of lowest order in the
coupling amplitude $\left \vert V_{\vec{k}}\right \vert ^{2}$. The main
difference of the present result is the incorporation of a finite momentum
exchange $\vec{k}$. This results in the following expression for the imaginary
part of the density response function (\ref{DensResp2}) which is proportional
to the Bragg response (\ref{BraggResp}):%
\begin{equation}
\operatorname{Im}\chi \left(  \omega,\vec{k}\right)  =\frac{1}{\omega^{4}}%
\frac{\pi}{m_{I}^{2}}\sum_{\vec{q}}\left(  \vec{k}.\vec{q}\right)
^{2}\left \vert V_{\vec{q}}\right \vert ^{2}S\left(  \vec{q}+\vec{k}%
,\omega-\omega_{\vec{q}}\right)  , \label{ImagDensResp}%
\end{equation}
where $S\left(  \vec{k},\omega \right)  $ is the dynamic structure factor of
the impurities:%
\begin{equation}
S\left(  \vec{k},\omega \right)  =\frac{1}{2\pi \hbar}\int_{-\infty}^{\infty
}dte^{i\omega t}\left \langle \psi \left \vert \rho_{\vec{k}}\left(  t\right)
\rho_{\vec{k}}^{\dag}\right \vert \psi \right \rangle . \label{DynStruct}%
\end{equation}

The response also allows a determination of an effective mass which we shall
call the spectral effective mass. This is done through an extension of the
f-sum rule that was introduced in Ref. \cite{PhysRevB.15.1212} and generalized
for the polaron in a condensate in Ref. \cite{PhysRevA.83.033631} to:%
\begin{equation}
\frac{N\pi}{2m^{\ast}}+\int_{0^{+}}^{\infty}d\omega \text{ }\omega \lim_{\vec
{k}\rightarrow0}\frac{\operatorname{Im}\left[  \chi \left(  \vec{k}%
,\omega \right)  \right]  }{k^{2}}=\frac{N\pi}{2m_{I}}. \label{somregel}%
\end{equation}
The origin of this sum rule is a $\delta$-peak at $\omega=0$ in the response,
the spectral weight of this $\delta$-peak in the $\vec{k}\rightarrow0$ limit
equals the first term in (\ref{somregel}). It is important to note that the
spectral effective mass as determined by (\ref{somregel}) is a single particle
property and is thus not the same as the inertial effective mass of expression
(\ref{EffMass}). Only in the limit of one polaron the two masses coincide as
can be checked from equation (\ref{somregel}). This polaronic spectral
effective mass, as defined in (\ref{somregel}), was for example measured in
Ref. \cite{PhysRevLett.100.226403} for the Fr\"{o}hlich solid state polaron.

Note that in this formalism the influence of multiple impurities on the
polaronic response is solely determined by the dynamic structure factor of the impurities.

\section{Spin-polarized fermionic impurity gas in a BEC}

In this section we apply the results of the previous section to the case of a
spin-polarized gas of fermionic impurities. Because of the Pauli exclusion
principle the s-wave scattering length is zero which results in essentially no
interactions between the impurities at low temperature. This means the
impurities can be described as an ideal fermionic gas for which the static and
dynamic structure factor are known. All results in this section are presented
in polaronic units, i.e. $\hbar=m_{I}=\xi=1$, and for lithium-6 impurities in
a sodium condensate, i.e. $m_{B}/m_{I}\approx3.82207$.

\subsection{Ground state properties}

The static structure factor of an ideal fermionic gas at zero temperature is
given by \cite{Pines}:%
\begin{equation}
S\left(  \vec{k}\right)  =\left \{
\begin{array}
[c]{cc}%
\frac{3}{2}\frac{k}{2k_{F}}-\frac{1}{2}\left(  \frac{k}{2k_{F}}\right)  ^{3} &
\operatorname*{if}\text{ }k<2k_{F}\\
1 & \operatorname*{if}\text{ }k\geq2k_{F}%
\end{array}
\right.  ,
\end{equation}
where $k_{F}$ is the Fermi wave vector which, for a non-degenerate gas, is
given by $k_{F}=\left(  6\pi^{2}n\right)  ^{1/3}$, with $n$ the impurity
density. Introducing the kinetic energy $\varepsilon_{kin}$ of an ideal
fermionic gas together with the Bogoliubov dispersion (\ref{BogDisp}) and the
interaction amplitude (\ref{IntAmp}) in the expression for the ground state
energy (\ref{GroundEnerg}) leads to:%
\begin{equation}
\frac{E}{N}=\frac{3}{10}k_{F}^{2}-\frac{\alpha}{2\pi}\left(  \frac{m_{B}%
+1}{m_{B}}\right)  ^{2}\int_{0}^{\infty}dkk^{2}\left[  \frac{m_{B}S^{2}\left(
\vec{k}\right)  }{k\sqrt{k^{2}+2}S\left(  \vec{k}\right)  +m_{B}k^{2}}\left(
\frac{k^{2}}{k^{2}+2}\right)  ^{1/2}-\frac{m_{B}}{m_{B}+1}\right]  ,
\label{EnVeelPol}%
\end{equation}%
\begin{figure}
[ptb]
\begin{center}
\includegraphics[
height=4.1846cm,
width=8.703cm
]%
{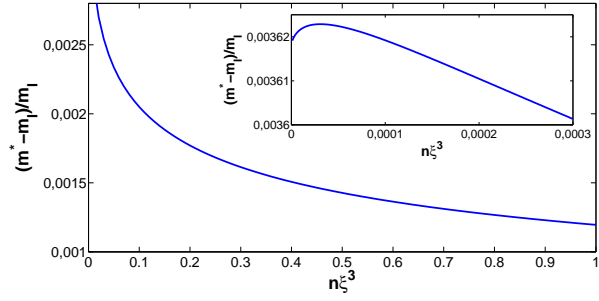}%
\caption{The inertial effective mass of the polarons as a function of the
density for lithium-6 impurities in a sodium condensate at $\alpha=0.01$
according to formula (\ref{EffMass2}). The inset shows the behavior at small
densities. In the limit $n\rightarrow0$ the effective mass of a single polaron
is retrieved and for $n\rightarrow \infty$ the effective mass becomes the bare
impurity mass. }%
\label{fig: EffMasGrond}%
\end{center}
\end{figure}
with $\alpha$ the polaronic coupling parameter from expression (\ref{CouplPar}%
). The first term represents the kinetic energy and the second is the
polaronic contribution, the interaction energy has vanished since we are
describing an ideal gas. Notice that in the polaronic contribution an
additional term has appeared which is needed to obtain a convergent energy and
was also obtained in Ref. \cite{PhysRevB.80.184504} as a result of the
renormalization of the boson-impurity interaction. The polaronic contribution
in equation (\ref{EnVeelPol}) grows linearly with $k_{F}$ as the number of
particles is increased. Since the kinetic energy of the Fermi gas grows as
$k_{F}^{2}$, the relative contribution of the polaronic energy with respect to
the kinetic energy decreases. In the limit of large densities, the kinetic
energy dominates the polaronic effect.

Introducing the Bogoliubov dispersion (\ref{BogDisp}) and the interaction
amplitude (\ref{IntAmp}) in the expression for the polaron inertial effective
mass (\ref{EffMass}) results in:%
\begin{equation}
m_{I}^{\ast}=1+\frac{4\alpha}{3\pi}\left(  \frac{m_{B}+1}{m_{B}}\right)
^{2}\int_{0}^{\infty}dk\frac{1}{\sqrt{2+k^{2}}}\frac{k^{2}S^{2}\left(  \vec
{k}\right)  }{\left(  \frac{1}{m_{B}}\sqrt{k^{2}+2}S\left(  \vec{k}\right)
+k\right)  ^{3}}. \label{EffMass2}%
\end{equation}
This expression is presented in figure \ref{fig: EffMasGrond} as a function of
the impurity density. In the limit $n\rightarrow0\ $the one-polaron result is
retrieved, which was already anticipated using formula (\ref{EffMass}). For
$n\rightarrow \infty$ the inertial effective mass becomes equal to the bare
impurity mass. This limit can easily be examined analytically with equation
(\ref{EffMass2}) which reveals a $n^{-1/3}$ behavior at large densities. For
the Fr\"{o}hlich solid state polaron the same qualitative behavior as in
figure \ref{fig: EffMasGrond} was found for the effective mass if the
electrons are described as a free gas\cite{PhysRevB.34.2621,
springerlink:10.1007/PL00011092}.

\subsection{Response to Bragg spectroscopy}

For the response the dynamic structure factor is needed, which for an ideal
fermionic gas at temperature zero is given by \cite{Mahan}:%
\begin{figure}
[ptb]
\begin{center}
\includegraphics[
height=4.1824cm,
width=8.7052cm
]%
{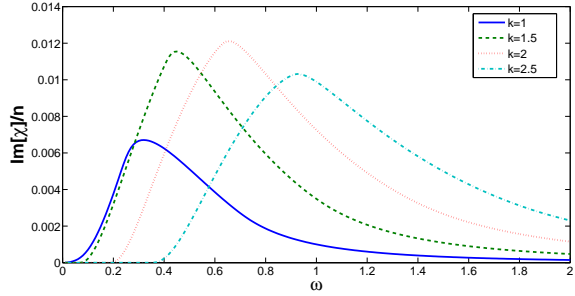}%
\caption{The Bragg response (\ref{ImDensResp}) of the polaronic system
consisting of polarized lithium-6 impurities in a sodium condensate as a
function of $\omega$ for different values of the exchanged momentum $k$. The
impurity density is taken $n=0.01$ and the polaronic coupling parameter is
$\alpha=0.01$. }%
\label{fig: RespKAfh0}%
\end{center}
\end{figure}
\begin{equation}
S\left(  \vec{q},\omega \right)  =\frac{1}{4\pi^{2}q}\theta \left(
\omega \right)  \left \{
\begin{array}
[c]{ccc}%
\omega & \text{if} & \frac{k_{F}^{2}}{2}>\frac{1}{2q^{2}}\left(  \omega
+\frac{q^{2}}{2}\right)  ^{2}\\
\frac{k_{F}^{2}}{2}-\frac{1}{2q^{2}}\left(  \omega-\frac{q^{2}}{2}\right)
^{2} & \text{if} & \frac{k_{F}^{2}}{2}<\frac{1}{2q^{2}}\left(  \omega
+\frac{q^{2}}{2}\right)  ^{2},\\
0 & \text{if} & \frac{k_{F}^{2}}{2}<\frac{1}{2q^{2}}\left(  \omega-\frac
{q^{2}}{2}\right)  ^{2}%
\end{array}
\right.
\end{equation}
with $\theta$ the Heaviside step function. Introducing the Bogoliubov
dispersion (\ref{BogDisp}) and the interaction amplitude (\ref{IntAmp}) in the
expression for the imaginary part of the density response function
(\ref{ImagDensResp}) leads to:%
\begin{equation}
\operatorname{Im}\chi \left(  \omega,\vec{k}\right)  =\frac{\alpha}{16\pi
\omega^{4}}\left(  \frac{m_{B}+1}{m_{B}}\right)  ^{2}\sum_{\vec{q}}\left(
\vec{k}.\vec{q}\right)  ^{2}\frac{q^{3}}{\sqrt{q^{2}+2}}S\left(  \vec{k}%
+\vec{q},\omega-\omega_{\vec{q}}\right)  . \label{ImDensResp}%
\end{equation}
In figure \ref{fig: RespKAfh0} expression (\ref{ImDensResp}) is shown as a
function of $\omega$ for various momentum exchanges. A peak is seen that
represents the emission of Bogoliubov excitations and which is shifted to
higher frequencies for larger momentum exchange. This behavior is to be
expected since more energy is needed to create a Bogoliubov excitation with a
higher momentum.

\begin{figure}
[ptb]
\begin{center}
\includegraphics[
height=4.1824cm,
width=8.7052cm
]%
{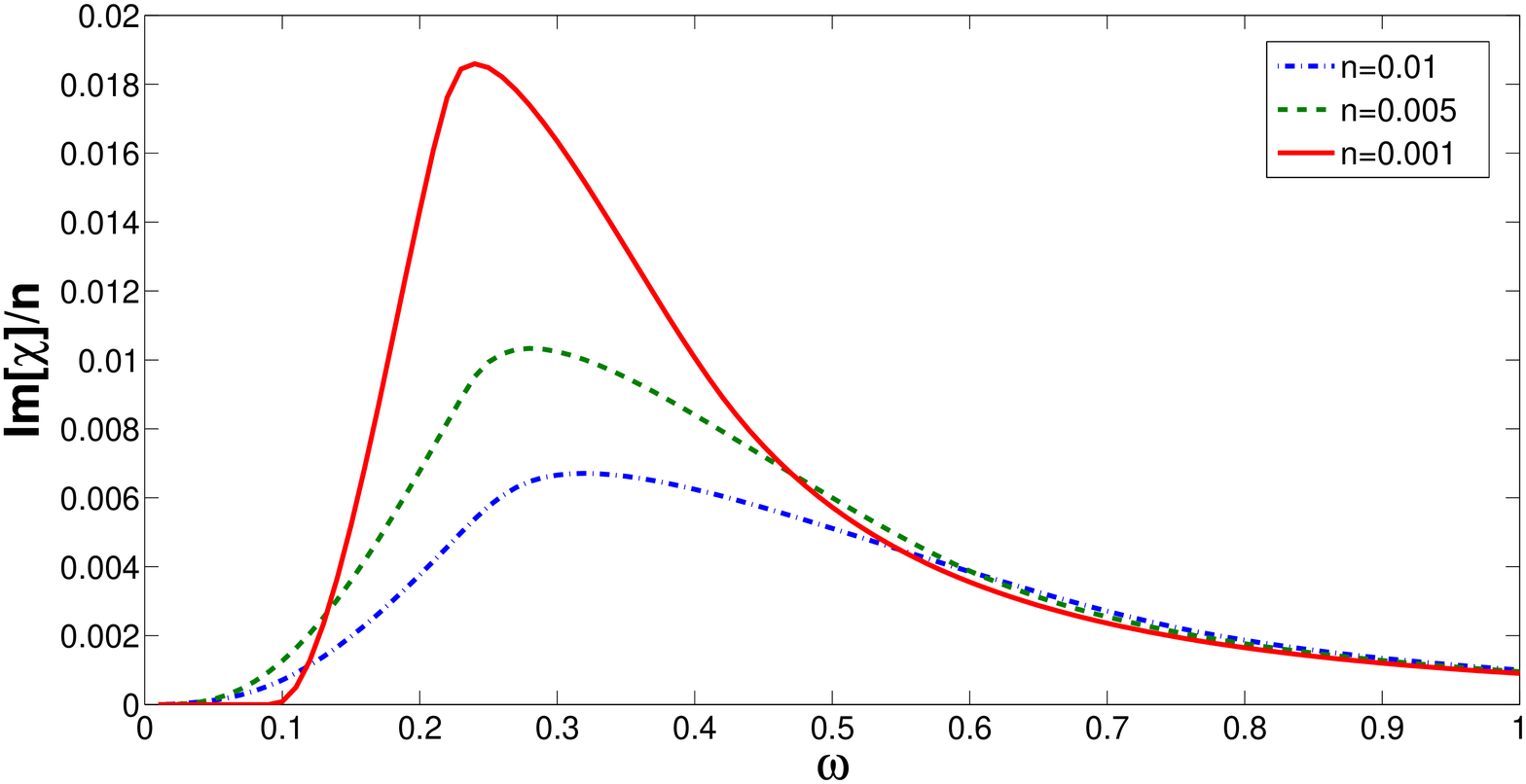}%
\caption{\ The Bragg response (\ref{ImDensResp}) of the polaronic system
consisting of polarized lithium-6 impurities in a sodium condensate as a
function of $\omega$ for different impurity densities $n$. The exchanged
momentum is taken $k=1$ and the polaronic coupling parameter is $\alpha=0.01$.
}%
\label{fig: RespDichtAfh0}%
\end{center}
\end{figure}

\begin{figure}
[ptb]
\begin{center}
\includegraphics[
height=4.1846cm,
width=8.7052cm
]%
{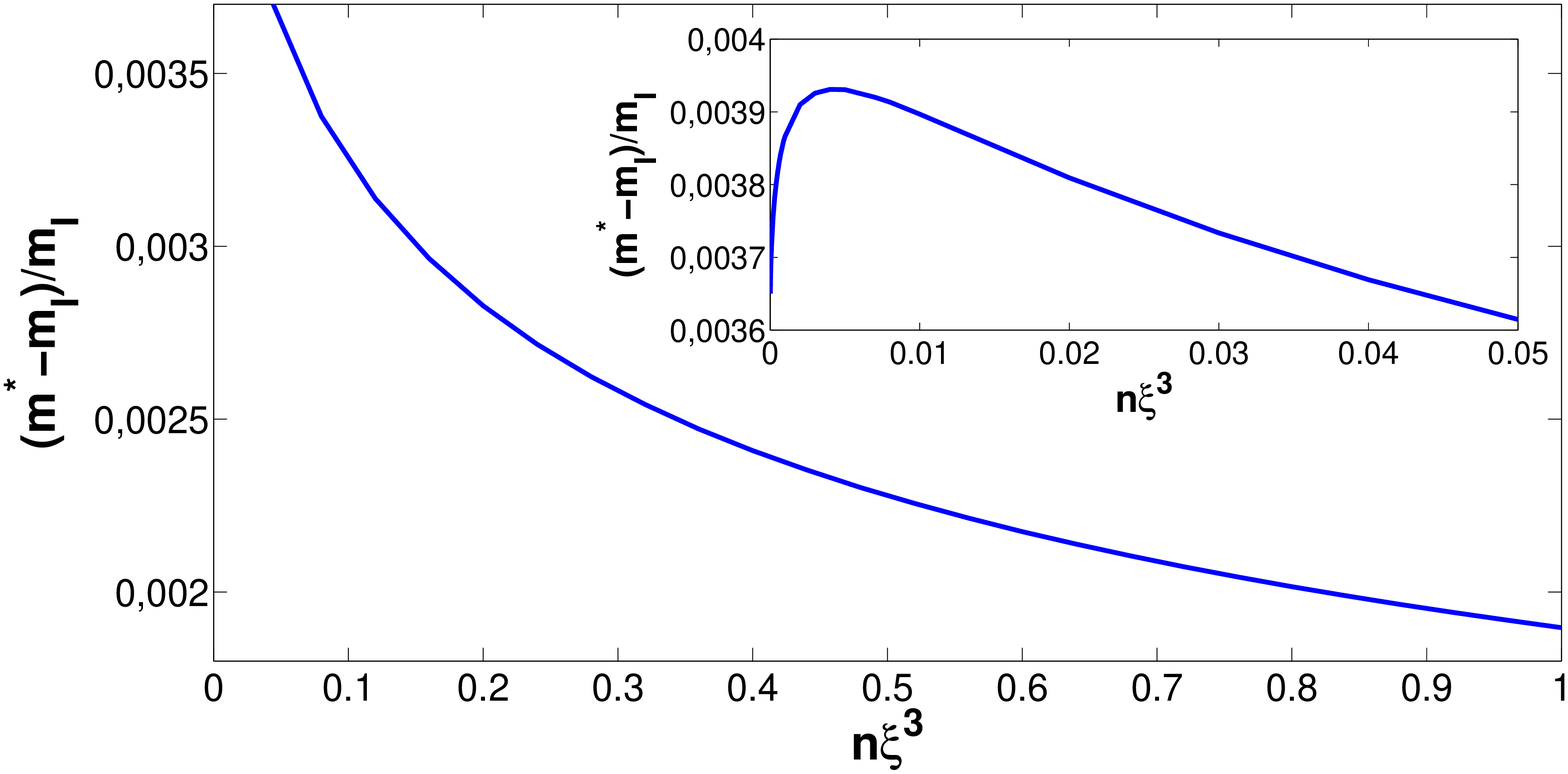}%
\caption{The spectral effective mass of the polarons as a function of the
impurity density for lithium-6 impurities in a sodium condensate at
$\alpha=0.01$ as determined by the sum rule (\ref{somregel}).}%
\label{fig: EffMassResp}%
\end{center}
\end{figure}

In figure \ref{fig: RespDichtAfh0} the peak in the Bragg response is presented
for different impurity densities. Note that for higher densities the spectral
weight of the $\omega>0$ peak diminishes. This is not in contrast with the
f-sum rule: the spectral weight of the $\omega=0$ delta peak in the spectrum
compensates, in accordance with expression (\ref{somregel}). For $\vec
{k}\rightarrow0$ the weight of this $\delta$-function is related to the
spectral effective mass such that an attenuation of the peak at $\omega>0$
corresponds to a decrease of the spectral effective mass. The spectral
effective mass deduced with the sum rule (\ref{somregel}) is presented in
figure \ref{fig: EffMassResp} as a function of the impurity density. Note that
the spectral effective mass behaves generally only qualitatively the same as
the inertial effective mass from figure \ref{fig: EffMasGrond} and only in the
limit $n\rightarrow0$ the same result is retrieved. This is because the
spectral effective mass is a one-particle property while the inertial
effective mass is related to the entire polaronic system.

We would like to emphasize that according to our calculations different
definitions of the effective mass result in another behavior and can in
general not be compared. This also means that for an experiment it is
important to know which effective mass is of importance for the specific
setup.%

\section{Conclusions}

The weak coupling many-polaron formalism which was developed in the context of
Fr\"{o}hlich solid state polarons in Refs. \cite{GroundstateManyPolaron} and
\cite{PhysRevB.64.104504} was applied in the present work to the case of the
polaronic system consisting of impurities in a Bose-Einstein condensate. The
properties of the ground state and the response to Bragg spectroscopy were
examined. The sum rule of Ref. \cite{PhysRevB.15.1212} which relates the
response to the spectral effective mass was formulated in the present context.
Also a calculation of the inertial effective mass was presented which is
related to the many-polaron system as a whole. It turns out that the ground
state properties are determined by the static structure factor of the
impurities while the response is governed by the dynamic structure factor, as
in the case of the Fr\"{o}hlich solid state polarons.

This generalization of the many-polaron formalism was then applied to the case
of spin-polarized fermionic impurities which behaves as an ideal gas for which
the structure factors are well-known. The numerical calculations were done for
lithium-6 impurities in a sodium condensate. Both the ground state properties
and the Bragg response indicate that in the limit of high impurity density the
polaron effect disappears, which is also the case for Fr\"{o}hlich solid state
polarons. In the Bragg response a peak was observed that corresponds to the
emission of Bogoliubov excitations. The behavior of this peak as a function of
the exchanged momentum and the impurity density was examined. It was also
shown that the two definitions for the effective mass exhibit a different
behavior, this is because the inertial effective mass is a property of the
system as a whole while the spectral effective mass is a one-particle property.

\bibliographystyle{apsrev}
\bibliography{acompat,beispiel,VeelPolaron}

\end{document}